\documentclass[pra,preprint,tightenlines,eqsecnum]{revtex4}

\begin{document}

 \title[Vacuum Energy and Heat Kernel]{Systematics of the 
Relationship between Vacuum Energy Calculations and Heat Kernel 
Coefficients} 

 \author{S. A. Fulling}
 \email{fulling@math.tamu.edu}
 \homepage{http://www.math.tamu.edu/~fulling}
\affiliation{Department of Mathematics, Texas A\&M University,%
  College Station, TX, 77843-3368 USA\\
{\rm and}\\
Mathematical Sciences Research Institute,%
Berkeley, CA, 94720-5070 USA}

\date{April 3 , 2003}

  \begin{abstract}
 Casimir energy is a nonlocal effect; its magnitude cannot be 
deduced from heat kernel expansions, even those including 
 the integrated boundary terms.
 On the other hand, it is known that the divergent terms in the 
regularized (but not yet renormalized) total vacuum energy are 
associated with the heat kernel coefficients.
Here a recent study of the relations among the eigenvalue density, 
the heat kernel, and the integral kernel of the operator 
 $e^{-t\sqrt{H}}$ is exploited to characterize this association 
completely.  Various previously isolated observations about the 
structure of the regularized energy emerge naturally. For over 20 
years controversies have persisted  stemming from the fact that 
 certain (presumably physically meaningful) terms in the renormalized 
 vacuum energy density in the interior of a cavity become singular 
at the boundary and correlate to certain divergent terms in the 
regularized total energy. 
 The point of view of the present paper promises to help resolve
 these issues. 
 \end{abstract} 

 \maketitle

 \section{Introduction}\label{sec:intro}

 The aim of this paper is to bring order and completeness into a 
welter of observations in the physics literature about various 
contributions to quantum vacuum energy (before renormalization) and 
their relations to the heat-kernel expansion. 
 The primary tool is 
the systematic theory of Riesz means of spectral densities and 
their relation to the asymptotic expansions of various integral 
kernels associated with the central partial differential operator 
of the field theory (e.g., [\onlinecite{riesz}]); 
 the Casimir energy,
regularized by an exponential ultraviolet cutoff, 
is identifiable with certain terms in these expansions \cite{naples}.
 Both total energy and local 
energy density are considered; for the latter, we concentrate on 
the singular asymptotic behavior near a boundary, 
 which was the 
subject of considerable calculational attention and physical 
controversy two decades ago (e.g., [\onlinecite{DC,KCD}]), 
 when  the 
Casimir effect was of great interest as a model of quantum effects 
in cosmology
(and in the bag model of hadrons). 
 Interest in Casimir energy has recently resurged 
because of greatly improved experiments \cite{Lamo,MR,RM,BCOR} and 
hints of technological application \cite{CAKBC}. 
 With it, the 
controversies have also reemerged (e.g., [\onlinecite{Milton,GJ4,Miltalk}]). 
(More complete bibliographical citations appear later in their 
proper contexts.)  

 Section \ref{sec:notat}
  sets up notation and summarizes the needed background 
information about the asymptotic expansions of integral kernels and 
the Riesz means of eigenvalue densities.
 Section~\ref{sec:model}
  concerns the energy density of a  scalar field, with 
emphasis on the instructive example of parallel plates with Robin 
boundary conditions;
 to avoid  confusion in the quantum theory later, it is important 
to note that the scalar field's energy contains, already at the 
classical level, a term concentrated on the boundary, unless the 
conformal-coupling parameter is given the nonstandard 
 value~$\frac14$. 
 The heart of the paper is Section~\ref{sec:schema},
  where the structures of the 
asymptotic expansions of the heat kernel and the total regularized 
energy (as functions of an auxiliary ``time'') and of the 
renormalized energy density (as function of the distance from the 
boundary) are presented and then correlated in dimensions $1$, $2$, 
and~$3$.
 In Section~\ref{sec:implic} we review the controversies over infinite 
boundary terms in the energy and draw some conclusions on the basis 
of the observations in the previous section.

  \section{Notation and framework}\label{sec:notat}

Let $H$ be a self-adjoint, elliptic, second-order differential operator,
independent of~$t$.
(To avoid some extraneous complications later, 
let us also assume that the spectrum of~$H$ is nonnegative.)
 Then the (hyperbolic) wave equation
 \begin{equation}
\frac{\partial^2\phi}{\partial t^2} =-H\phi \label{wave} 
\end{equation}
 defines a field theory with time-translation invariance. 
 Here we consider only scalar fields, but the extension to, for instance,
the electromagnetic field would be straightforward.
 Also, we are primarily concerned here with cavities in flat space, 
without potentials, so that $H$ is just the negative of the 
Laplacian equipped with certain boundary conditions; 
 however, most of what follows applies with only minor 
complications to more general operators \[H = - g(x)^{\mu\nu} 
\nabla^A_\mu \nabla^A_\nu +V(x)\] 
 with gravitational, gauge, and scalar potentials.

For present purposes, we take the {\em regularized vacuum energy\/} 
of the quantum field to be {\em defined\/} as 
 \begin{equation} 
E(t) \equiv \frac12 \sum_n \omega_n e^{-\omega_n t},  
\label{energy}
\end{equation}
where $\omega_n{}\!^2$ are the eigenvalues of~$H$, and $t$ is a 
parameter with dimensions of time --- not the physical time in 
(\ref{wave}). 
 (Of course, at this point one is assuming that $H$ has only 
discrete spectrum. 
 Continuous spectrum is normally associated with an unbounded 
region in space, for which one would not expect a finite total 
energy to exist.) 
 The physical (renormalized) energy is identified 
with the constant term in the asymptotic expansion of $E(t)$ as 
$t\to 0$, in keeping with the expectation that each mode of the 
field with frequency $\omega$ should contribute energy $\frac12 
\omega$. 
 Formula (\ref{energy}) is a standard, but admittedly 
arbitrary, and recently somewhat unfashionable, way of regularizing 
the energy. 
 The relation of this and other ``ultraviolet'' or ``cutoff'' 
regularizations to the now more popular ``analytic'' methods 
(notably dimensional regularization and zeta functions) has been 
clarified by several authors [\onlinecite{CVZ,Kay,SS1,SS2,SS3,BS,Sant}]. 

In quantum field theory one has not only a total energy but also a 
local energy density, $T_{00}(x)$ (and, indeed, an energy-momentum 
(stress) tensor, $T_{\mu\nu}(x)$). 
 The energy density remains meaningful in unbounded regions, where 
the total energy is not defined (and $H$ may have continuous 
spectrum). 
 There is a formal expression for $T_{00}(x)$ as a sum or integral 
over the spectrum, the summand being built out of the 
eigenfunctions, or the spectral projection kernel, and their 
derivatives at~$x$; for the details we refer to the literature 
(e.g.,~[\onlinecite{BD}]) and Sec.~\ref{sec:model}. 
 From that expression a regularized quantity  $T_{00}(x;t)$ is 
defined by inserting an exponential cutoff in precise analogy 
with~(\ref{energy}). 
 In fact, it is known that this approach is not quite adequate when 
spatial curvature, or indeed any potential, is present, because 
{\em trace anomalies\/} arise; the conceptually closest adequate 
approach is then {\em covariant point splitting\/}      (e.g., 
[\onlinecite{BCF}]). 
 That complication is irrelevant, however, to cavities in flat 
space without potentials, which are our concern at present. 

Let $d$ be the spatial dimension,  
let $M$ denote the $d$-dimensional region in question (where $H$ acts), 
and let $t$ be an 
auxiliary variable ranging from $0$ to $+\infty$. 
 The well known {\em heat kernel},
$K(t,x,y) \equiv \langle x|e^{-tH}|y\rangle$,
solves the heat equation associated with~$H$, in the sense that
\[ u(t,x) \equiv \int_M K(t,x,y) f(y)\,dy \]
is the unique solution of the (parabolic) initial-value problem 
\[ \frac{\partial u}{\partial t} = -Hu, \quad u(0,x) = f(x). \]
It is well known that $K$ possesses an asymptotic expansion of the form
\begin{equation}
K(t) \sim \sum_{s=0}^\infty b_s t^{-\frac d2 + \frac s2}. 
\label{heat} \end{equation}
More precisely: 
 (a) If $x$ is a point in the interior of $M$, then the diagonal 
value $K(t,x,x)$ has an expansion of form (\ref{heat}), where 
$b_s$ depends on~$x$. 
 (b) When $M$ is bounded,  the ``trace'' $\int_M K(t,x,x)\, dx$ has 
an expansion of form (\ref{heat}). 
 (The second statement is not a 
trivial consequence of the first, because expansion (a) is 
nonuniform in distance from the boundary.  
 In fact, for a flat-space cavity $b_s(x)$ in (a) is zero for all  
$s>0$, but the same is not true for the number $b_s$ in (b).) 
 For a 
cavity, $b_0$ in (b) is proportional to the volume of~$M$, and the 
higher coefficients are integrals over the boundary of~$M$ whose 
details depend on the boundary conditions and on the curvature of 
the boundary. 
 For a more general~$H$, similar local integral 
formulas for $b_s$ in terms of the potentials, curvature, etc.\ 
inside~$M$ and on its boundary are known in great detail for small 
values of~$s$, their calculation for ever-increasing~$s$ being a 
major mathematical industry (e.g., [\onlinecite{Kirsten}]). 

The less well known {\em cylinder kernel},
$T(t,x,y) \equiv \langle x|e^{-t\sqrt{H}}|y\rangle$,
can be defined similarly:
\[ u(t,x) \equiv \int_M T(t,x,y) f(y)\,dy \]
is the unique bounded solution of the (elliptic) boundary-value problem 
\[ \frac{\partial^2 u}{\partial t^2} = +Hu, \quad u(0,x) = f(x). \]
(The region in $(t,x)$-space is a semiinfinite cylinder with base~$M$.)
The counterpart of (\ref{heat}) is
\begin{equation}
T(t) \sim \sum_{s=0}^\infty e_s t^{- d +  s}
+ \sum_{\scriptstyle s=d+1 \atop
 \scriptstyle s-d \mbox{{} \scriptsize odd}}^\infty 
f_s t^{-d+s} \ln t. 
\label{cyl} \end{equation}
(Again, (\ref{cyl}) has both (a)- and (b)-type interpretations.)
The central fact relating these expansions 
(which in essence appears in [\onlinecite[Sec.~III]{CVZ}]) is:

\bigskip
\noindent{\bf Theorem.} 
\begin{enumerate}
\item If $s-d$ is even or negative,
\begin{equation}
e_s = \pi^{-1/2} 2^{d-s}
  \Gamma\left( \frac d2 - \frac s2 + \frac12 \right) b_s
\label{eb} \end{equation}
(and $f_s=0$).
\item If $s-d$ is odd and positive,
\begin{equation}
f_s = (-1)^{(s-d+1)/2} \pi^{-1/2} 2^{d-s+1} 
\frac1{\Gamma\left( \frac s2 - \frac d2 +\frac12 \right)} \, b_s \,,
\label{fb} \end{equation}
whereas $e_s$ is undetermined by the heat kernel expansion.
\end{enumerate}
\bigskip

Before explaining where this theorem comes from, 
 we make some remarks upon it.
 First, in (\ref{eb}) and (\ref{fb}) the gamma functions 
have no zeros or poles for the relevant values of the parameters; 
therefore, a zero or nonzero $b_s$ is always associated, 
respectively, with a zero or nonzero value for the coefficient on 
the left-hand side of the equation. 
 Second, the ``new'' coefficients $e_s$ ($s-d$ odd and positive) 
are inherently {\em global\/} in their dependence on the geometry 
of~$M$: 
 There are no local integral formulas like those for~$b_s$ (and 
hence for the left-hand sides of (\ref{eb}) and 
 (\ref{fb})) \cite{GG}. 
  Third (see Sec.~\ref{sec:schema}),
  the renormalized vacuum energy is equal to $-\frac12 e_{1+d}$ 
(the first of the new, nonlocal coefficients) 
from expansion (b), plus possible local terms.
 (Note that a purely local formula for vacuum energy would imply 
that the energy associated with parallel plates is independent of 
their separation, hence that there is no Casimir force!) 
 Fourth, for a particular $H$ any or all of the logarithmic coefficients 
$f_s$ may happen to vanish, but the nonlocal, independent nature of the 
corresponding $e_s$ persists in such a case.

 The theorem can be proved by relating the two expansions, 
(\ref{heat}) and (\ref{cyl}),
 to certain ``moments'' 
(in a generalized sense of the word)
characterizing the asymptotic behavior of 
the eigenvalue density (or of a local spectral density, 
 the inverse Laplace transform of $K(t,x,x)$, in the   
``untraced'' (a) situation).
 One suitable set of such quantities consists of residues and 
values, at certain points,
  of the zeta functions associated with the operators $H$ and 
$\sqrt H$.
 In that case the proof is an adaptation and generalization of the 
proof of Lemma 2.2 of~[\onlinecite{Gil4}].
 (See also~[\onlinecite{CVZ}].)
 We omit the details, citing only the crucial equation
\begin{equation}
\zeta_{\sqrt{H}}(2s) = \zeta_H(s).
\label{zeta}\end{equation}

 An alternative set, with more direct spectral and physical 
significance, consists of the {\em Riesz means\/} of the eigenvalue 
density.
 These are the rigorous counterpart of the formal high-frequency 
asymptotic expansions of the eigenvalue density whose term-by-term 
Laplace transforms yield the expansions (\ref{heat}) and~(\ref{cyl}).
 The importance of Riesz means in understanding (\ref{cyl}), and 
vice versa, has been developed in [\onlinecite{leipzig,riesz,FGR}].  
(Papers [\onlinecite{E,EF,EGV}] are related.)
 The present paper demonstrates the usefulness of that work in 
physics.
 We shall review it here, using the language appropriate to the 
``traced'' (b) situation.

 Riesz means \cite{Har,Hor1,Hor2} generalize the Ces\`aro means 
used in place of partial sums to accelerate the convergence of 
Fourier series. 
Let $\mu(\lambda)$ be the number of eigenvalues of $H$
 less than~$\lambda$
 (the counting function, or ``spectral staircase'').
 Write
 \begin{equation}
 \lambda=\omega^2 \qquad(\omega>0). \label{omega}
 \end{equation}
 (Thus, if $\lambda$ is an eigenvalue,
  $\omega$ is the frequency of the normal mode(s) of the wave 
equation associated with~$\lambda$;
 also, $\omega$ is the corresponding eigenvalue of~$\sqrt H$.)
 For a positive integer~$\alpha$ the Riesz mean 
$R^\alpha_\lambda\mu(\lambda)$
 is defined by averaging $\mu$ over an $\alpha$-dimensional 
simplex:
 \begin{equation}
R^\alpha_\lambda\mu(\lambda) \equiv
 \frac1{\alpha!}\,\lambda^{-\alpha} \int^\lambda 
{\buildrel\alpha\over\cdots}
 \int \mu(\tilde{\lambda})\,d\tilde\lambda,
 \label{Rlambda} \end{equation}
 where $\alpha$ successive indefinite integrations  are implied.
 (This operation can be reexpressed as a single integral with a 
nontrivial kernel function.)
 A different sequence of Riesz means is defined by integration 
over~$\omega$:
 \begin{equation}
R^\alpha_\omega\mu(\omega) \equiv
 \frac1{\alpha!}\,\omega^{-\alpha} \int^\omega 
{\buildrel\alpha\over\cdots}
 \int \mu(\tilde{\omega}^2)\,d\tilde\omega,
 \label{Romega} \end{equation}

Starting from the known properties of the heat-kernel expansion 
(\ref{heat}),
 one can prove the finite asymptotic approximations
 \begin{equation}
R^\alpha_\lambda\mu = \sum_{s=0}^\alpha a_{\alpha s} \lambda^{d-s\over2}
 + O\left(\lambda^{d-\alpha-1\over2}\right),
 \label{Rlam}\end{equation}
 \begin{equation}
R^\alpha_\omega\mu = \sum_{s=0}^\alpha c_{\alpha s} \omega^{d-s}
 +\sum^\alpha_{\scriptstyle s=d+1\atop \scriptstyle s-d 
 \mbox{\scriptsize{} odd}} d_{\alpha s} \omega^{d-s} \ln \omega
 + O\left(\omega^{d-\alpha-1}\ln \omega\right).
 \label{Rome}\end{equation}
Here the $a$ coefficients are related quite directly to the 
 heat-kernel coefficients:
 \begin{equation}
 b_s = {\Gamma\left({d+s\over2} +1 \right) \over
 \Gamma(s+1)} \, a_{ss}\,.
 \label{ba}\end{equation}
 (The $a_{\alpha s}$ with $\alpha \ne s$ have similar formulas, 
which we ignore here because those coefficients contain no 
additional information.)
Similarly, the $c$ and $d$ coefficients are related to the   
cylinder expansion by
[$\psi\equiv (\ln\Gamma)'$]
 \begin{equation}
 e_s = {\Gamma(d+1)\over \Gamma(s+1)}\, c_{ss}
 \quad\mbox{if $d-s$ is even or positive},
 \label{ec}\end{equation}
 \begin{equation}
 f_s = -\, {\Gamma(d+1)\over \Gamma(s+1)}\, d_{ss}\,, \quad
e_s = {\Gamma(d+1)\over \Gamma(s+1)} [c_{ss}+ \psi(d+1)d_{ss}]
 \quad
 \mbox{if $d-s$ is odd and negative}. 
 \label{fedc}\end{equation}

The  final link is the connection between $(c_{ss},d_{ss})$ and 
$a_{ss}\,$.
 In [\onlinecite{riesz}] the ratio of $c_{ss}$ to $a_{ss}$ was given 
as a complicated finite sum of gamma functions.
 However, hidden in the appendix of [\onlinecite{riesz}]  
(because the authors did not appreciate their significance at the 
time)
are formulas that, after some routine manipulation, yield the 
relatively simple result
 \begin{equation}
 c_{ss} =  {\Gamma\left(\frac d2 -\frac s2 +\frac12\right) 
 \Gamma\left(\frac d2 +\frac s2 +1\right) 
  \over 2^s \,\Gamma\left(\frac d2 + \frac12\right) \Gamma\left(\frac 
d2 +1\right)}
 \, a_{ss} \quad\mbox{if $d-s$ is even or positive}.
 \label{ca}\end{equation}
 When $d-s$ is odd and negative 
(when the first factor in the numerator of (\ref{ca}) has a pole),
 [\onlinecite{riesz}] shows that
 {\em $c_{ss}$ is undetermined by $a_{ss}$}
 while $d_{ss}$ comes into existence:
 \begin{equation}
 d_{ss} = {(-1)^{d+1} \over (s-d-1)!\, d!} \,
 {\Gamma\left(\frac s2 -\frac d2\right) \over
 \Gamma\left(-\frac d2 -\frac s2\right)} \, a_{ss}
\quad\mbox{if $d-s$ is odd and negative}.
 \label{da}\end{equation}

 Equations (\ref{eb}) and (\ref{fb})
 now follow from (\ref{ba})--(\ref{da}) with aid of
\begin{equation}
\Gamma(z)\Gamma\left(z+{\textstyle\frac12}\right)
=\sqrt{\pi}\, 2^{1-2z}\,\Gamma(2z), \qquad
\Gamma(-z)\Gamma(z+1) ={-\pi \over \sin {\pi z}}\,.
\label{gamma}\end{equation}

Note that the Riesz means (\ref{Rlambda}) and (\ref{Romega})  
provide two families of definitions of an ``averaged density of 
states'' with  certain weightings. 
 More precisely, $R_\lambda^\alpha\mu$ or $R_\omega^\alpha\mu$ 
somehow describes the {\em averaged asymptotic behavior\/} of the 
eigenvalue density $\mu'(\lambda)$ at high frequency, but it also 
contains {\em constants of integration\/} summarizing what happens to 
$\mu'(\lambda)$ at low frequency. 
 (As $\alpha$ increases, the smoothing out of the eigenvalue 
distribution becomes more drastic, so that the expansion 
(\ref{Rlam}) or (\ref{Rome}) can be meaningfully  extended to a 
higher order; but at the same time, the contribution from the lower 
frequencies becomes more dominant.) 
 The crucial fact is that the  coefficients in the omega-mean 
expansion 
 (equivalently, those in the cylinder-kernel expansion)
 contain more information than the coefficients in the lambda-mean 
(or the heat-kernel) expansion. 
 Specifically, $c_{ss}$ for $s-d$ odd and positive is an 
independent, nonlocal spectral invariant. 
 In particular, $s-d=1$ corresponds (modulo local terms) to the 
Casimir vacuum energy.

  \section{Energy density, Robin boundary 
  conditions, and parallel plates}\label{sec:model} 

The energy density of a scalar field in flat space-time is 
(e.g., [\onlinecite{BD}])
 \begin{equation}
 T_{00}= \frac12\biggl[\left(\frac{\partial\phi}{\partial 
t}\right)^2
 + (1-4\xi)(\nabla \phi)^2 -4\xi \phi \nabla^2\phi \biggr],
 \label{T00}\end{equation}
where $\xi$ is an arbitrary number called the 
 {\em conformal coupling constant}.
 In curved space-time $\xi$ derives from a term 
 (in the Lagrangian or the equation of motion)
 coupling $\phi$ to the  curvature; 
 in a gravitational context, therefore, different 
values of $\xi$ correspond to different physical theories.
 In flat space, however, $\xi$ does not appear in the equation of 
motion, nor in the spectral decomposition of~$H$, 
 and the terms proportional to $\xi$ in (\ref{T00}) form a 
total divergence, $-2\xi\nabla \cdot (\phi \nabla \phi)$.
 The facile conclusion is that all values of $\xi$ yield the same 
total energy and hence are physically equivalent.
 We shall see, however, that subtleties arise at both classical and 
quantum levels.

 Because it leads to nontrivial phenomena even in one dimension,
 it is instructive to study the {\em Robin boundary 
condition},
 \begin{equation} \frac{\partial\phi}{\partial n} = \gamma \phi.
 \label{robin}\end{equation}
Here $\gamma(x)$ is, in general, a function defined on the boundary 
of the region~$M$, 
 and $\frac{\partial}{\partial n}$ is
 the outward normal derivative.
Dimensionally, $\gamma$~is an inverse length.
Romeo and Saharian \cite{RS}  have recently made an exhaustive study
 of the Casimir-like situation of two parallel plates with a 
constant value of $\gamma$ on each plate.
 (A continuing series of papers by Saharian and coworkers 
address similar issues in more complicated geometries).
Of course, the special cases of the Neumann 
($\gamma=0$) and Dirichlet (formally, $\gamma=\infty$) boundary 
conditions have been understood for a long time. 
With the sign convention in (\ref{robin}), 
our condition of nonnegative spectrum for $H$ 
is guaranteed by taking $\gamma$ negative;
Romeo and Saharian consider both signs of $\gamma$ 
(which they call $\beta^{-1}$), ignoring  the contributions 
to the energy from modes with negative eigenvalues, 
which do not affect the renormalization theory.

If $M$ is bounded, one should be able to integrate $T_{00}$ over 
$M$ to obtain a conserved energy.  (For parallel plates, 
integration in the perpendicular dimension yields an energy per 
unit plate area.)
 In fact, in the Robin theory it is necessary to include in the 
energy a surface term concentrated on the boundary:
 \begin{equation}
 E\equiv \int_M T_{00}(x) \, d^d x + \frac12(4\xi-1)\int_{\partial 
M} \gamma \phi^2\, dS
 \label{totalE}\end{equation}
 (where $dS$ is the element of $(d-1)$-dimensional surface area on 
the boundary, $\partial M$).
 A calculation, using the equation of motion (\ref{wave}) 
for~$\phi$ and an integration by  parts,
 shows that $\frac{\partial E}{\partial t} =0$,
 the boundary term being essential for this result if $\gamma\ne 0$ 
and $\xi \ne \frac 14$.
 Such a term was apparently introduced into quantum field 
theory --- through a different rationale --- in [\onlinecite{KCD}].
Classically, the term is presumably related to the fact that a 
 one-dimensional Robin problem models a vibrating string attached
 to a point mass, which can exchange energy with the string
 (e.g., [\onlinecite[p.~15]{CZ}]).

 Three things should be noted about the boundary term in 
(\ref{totalE}).
 First, it is a completely classical matter;
 in particular, it is not a renormalization counterterm (at least, 
not yet).
 Second, it arises only in the full Robin theory:
 it vanishes in the pure Neumann case ($\gamma=0$) and the 
Dirichlet case ($\phi=0$, where the conservation of $E$ without 
that term can be easily verified).
 Third, even in the Robin case it vanishes if $\xi=\frac14$.
 Changing $\xi$ (in flat space) amounts to redefining some of the 
boundary energy as spread throughout the interior of~$M$ by 
integration by parts, and $\frac14$ is the choice that moves the 
energy {\sl entirely\/} into the interior.
 The striking thing about this value of $\xi$ is that it is neither 
the ``minimal'' choice ($\xi=0$) nor
 the choice that makes the full theory, including gravitation,
 conformally invariant.
 The ``conformal'' value is well known to be
 \begin{equation}
 \xi_{\rm c} \equiv {d-1\over 4d}\,,
 \label{conformal}\end{equation}
 which equals $0$ when $d=1$, equals $\frac16$ when $d=3$, and 
approaches $\frac14$ as $d\to\infty$.

In the quantum field theory one evaluates the expectation values
 of observables such as $T_{00}(x)$ and $E$ in the vacuum, or 
ground state,  of the field.
 The general rule of thumb~\cite{BD} for doing so, when the 
observable is classically defined by a quadratic form 
$T(\phi,\phi)$, 
 is to evaluate the associated bilinear form on  the normal modes 
of the field and sum:
 \begin{equation}
  \langle0|T|0\rangle= \sum_n T(\tilde\phi_n, \tilde\phi_n^*). 
 \label{bilinear}\end{equation}
 Here $\tilde\phi(t,x) = (2\omega_n)^{-1/2} \phi_n(x)e^{-i\omega_n 
t}$, where $\phi_n$ is an eigenvector of $H$ with eigenvalue 
$\omega_n{}\!^2$;
 generalization to continuous spectrum is possible.
(Henceforth the vacuum bra-ket symbols will be omitted 
when there is no danger of confusion.)
 The result of this formal calculation is usually a divergent sum 
or integral, which must be regularized --- perhaps by inserting
 an exponential ultraviolet cutoff, exactly as in (\ref{energy})
  --- and ultimately renormalized.
 (However, the calculations of energy density in [\onlinecite{RS}] 
do not require an explicit regularization step; instead, the 
renormalization is accomplished by isolating and subtracting at the 
integrand stage the terms corresponding to the vacuum state in 
infinite flat space.) 
 When such calculations were done some 20--30 years ago for the 
Dirichlet and Neumann problems with parallel plates,
 it was found that when $\xi=\xi_{\rm c}$
 the (renormalized) energy density $T_{00}(x)$ is {\em constant\/} 
(and nonzero) everywhere between the plates, thus yielding a finite 
Casimir energy (per unit area) that must be attributed to the 
global configuration of the plates, not to an interaction between 
the field and the boundary per se. 
 (In the Neumann case the contribution of the mode with $\omega=0$ 
is being ignored here.
When a zero-frequency mode exists, there is no true vacuum state, 
but vacuum conditions can be approached arbitrarily closely.)
 On the other hand, for $\xi\ne\xi_{\rm c}$
 the renormalized energy density includes a contribution strongly 
concentrated near the plates themselves; 
that part is, in fact, nonintegrable, hence at least superficially 
inconsistent  with  a finite value for the renormalized total 
energy.
 When the boundary is curved,    there are such near-boundary terms 
even when the coupling constant takes the conformal value, although 
the singularity at the boundary is weaker in that case.
 The recent Robin results \cite{RS} are a consistent extension of 
that picture:
 
 \begin{itemize}
 \item For a single (flat) plate, the renormalized $T_{00}$ vanishes 
if $\xi=\xi_{\rm c}$, regardless of~$\gamma$.

 \item In general, the energy density around a single plate is 
proportional to $\xi-\xi_{\rm c}$.
 If $x$ is distance from the plate, $T_{00}(x)$ is proportional to 
$x^{-(d+1)}$ in the Dirichlet and Neumann cases, and in the general 
Robin case it has additional terms proportional to  
 $\gamma^k x^{k-(d+1)}$ ($k=1,\ldots,d$)
 and $\ln |\gamma x|$.
(Romeo and Saharian actually find an exact expression for $T_{00}$, 
of which these terms are merely the divergent part of the 
asymptotic expansion for small~$x$.)

 \item For two parallel plates, $T_{00}$ is constant in the space 
between the plates if $\xi=\xi_{\rm c}$, with a rather 
complicated dependence on the parameters $a\gamma_1$ and 
$a\gamma_2$
($a$ being the plate separation and  $\gamma_i$ the Robin 
coefficients on the two plates),
 which reduces for the pure Dirichlet or Neumann problem to the 
known $a^{-(d+1)}$ dependence.

\item For parallel plates with $\xi\ne\xi_{\rm c}$,
 the result is even more complicated, but it can be written as the 
sum of the effects of the individual plates (described above) plus 
an interaction term that is everywhere finite.
 
 \item Romeo and Saharian perform an independent calculation of the 
renormalized total energy by the zeta-function method.
 This calculation is independent of $\xi$, as the eigenvalues do 
not depend on~$\xi$.
 The result includes a term that depends logarithmically on an 
arbitrary scale parameter --- indicating that a logarithmic 
divergence would appear in some other regularization methods.
 This term is independent of $a$, hence presumably unobservable.

 \item The authors demonstrate a certain formal consistency between 
 their energy-density and total-energy calculations.
 In particular, for a single plate they find 
 \begin{equation}
 E_{\rm vol} \equiv \int_0^\infty T_{00}(x)\,dx
  =4d(\xi-\xi_{\rm c})E,
 \quad E_{\rm surf} \equiv
\frac12(4\xi-1) \gamma \langle0|\phi(0)^2|0\rangle = d (1-4\xi)E,
 \label{Edecomp}\end{equation}
 where $E=E_{\rm vol}+E_{\rm surf}$
 is the putative total energy per unit area (on one side) of the plate.
 However, they point out that the integrals defining the three 
quantities in (\ref{Edecomp}) are actually divergent.
 One can eliminate the volume term by taking 
$\xi=\xi_{\rm c}\,$, or one can eliminate the surface 
term by taking $\xi=\frac14\,$, but not both at once.

 \end{itemize}

 In view of the last point, a strong case can be made that the most 
convenient choice of $\xi$ is $\frac14\,$, rather than the 
traditional values, $0$ or $\xi_{\rm c}\,$.
 One can then avoid accounting for and renormalizing the surface 
term, and one can be confident that any volume contribution must be 
taken seriously, not dismissed as an artifact.

 We have summarized the paper [\onlinecite{RS}] at some length 
because the parallel-plate Robin system is such an excellent, 
explicitly solvable model of the phenomena and issues that arise in 
more general systems, notably cavities with curved boundaries.
 The principal extrinsic curvatures of a boundary surface have the 
dimension of inverse length and enter calculations or expansions of 
energy density and total energy in very much the same way as does 
the Robin coefficient~$\gamma$.
 The leading divergence in $T_{00}$ as the boundary is approached 
is ubiquitous when $\xi$ does not have its conformal value;
 unlike the Dirichlet case, the classical surface term in the total 
energy of the Robin system interferes with the facile dismissal of 
this problem as some kind of artifact of a poor definition of 
energy density
 (whereas the nonexistence of the surface term in the Dirichlet 
case shows that the problem can't be resolved by a simple 
cancellation of volume and surface contributions, either).
 The higher-order terms in the expansion of $T_{00}$ near the 
boundary, which appear for any $\xi$ when $\gamma\ne0$,
  are closely analogous to terms that appear near curved 
boundaries for the electromagnetic field, as well as the conformally 
coupled scalar one.

   \section{Schematics of the expansions}\label{sec:schema}

Consider now a general cavity whose boundary is characterized by 
an extrinsic curvature tensor, $\kappa(x)$, and also equipped with a Robin
function, $\gamma(x)$.
The structure of the  heat expansion (\ref{heat}), for the trace, is
\begin{equation} 
K(t) \sim t^{-d/2} \bigl[
V + St^{1/2} + (\kappa + \gamma)t +
(\kappa^2 + \gamma^2 + \kappa\gamma)t^{3/2}
+ (\kappa^3 + \cdots)t^2 + \cdots\,\bigr].
\label{K}\end{equation}
Here, {\em and in all similar equations later},
  all the terms are indicated very schematically:
{\em all\/} numerical coefficients, 
 the integrations over $\partial M$, and the 
tensorial nature of $\kappa$ when $d\ge3$ are suppressed,
 as are terms involving derivatives of $\gamma$ or~$\kappa$
 (which are not important in low orders and dimensions).
 $V$ is the volume of $M$, and $S$ is the surface area of $M$
(more precisely, the $(d-1)$-dimensional volume of $\partial M$).

 The schematic  form of the (traced) 
 cylinder expansion (\ref{cyl}) now follows by the 
theorem in Sec.~\ref{sec:notat}.
 The  significance of the structure depends on the dimension.
 In the following formulas, the coefficients in boldface are the 
new, nonlocal terms, $e_{s}\,$,
 and those in square brackets are constant terms, which do not contribute 
to the regularized energy,
 \begin{equation}
 E(t) = - \frac12 \, {d T \over d t}\,.
 \label{Ereg}\end{equation}
     (The trace of the operator $e^{-t\sqrt{H}}$ is $\sum_n e^{-
     \omega_n t}$, so (\ref{Ereg}) is equivalent to 
     (\ref{energy}).)
  \begin{equation}
\hfilneg d=1:\qquad\hfill T(t) \sim Vt^{-1} + [S] 
 + \gamma t\ln t + {\bf E}t
 +\gamma^2t^2 + \gamma^3 t^3 \ln t + {\bf F}t^3 + \cdots\,.
\hfill \label{T1}\end{equation}
(In dimension $1$ there is no curvature, and the ``surface area'' 
$S$ degenerates to the number of endpoints of the region --- 
usually~$2$.  The $\gamma$ terms are also sums over the endpoints.)
 \begin{equation}
 \hfilneg d=2:\qquad\hfill
 T(t)\sim Vt^{-2} + St^{-1} + [\kappa+\gamma]
 + (\kappa^2+\gamma^2+\kappa\gamma) t\ln t
 +{\bf E}t + (\kappa^3 +\cdots\,)t^2 + \cdots\,.
 \hfill\label{T2}\end{equation}
 \begin{equation}
\hfilneg d=3: \qquad\hfill
 T(t) \sim Vt^{-3} +St^{-2} + (\kappa+\gamma)t^{-1} 
 + [\kappa^2+\gamma^2+\kappa\gamma] + (\kappa^3+\cdots\,) t\ln t
+ {\bf E}t +\cdots\,.
\hfill \label{T3}\end{equation}

 The untraced kernel expansions can be discussed similarly, but 
there is less to say.
 In the cavity setting, where there are no potentials or spatial 
curvature, the local heat expansion is trivial:
 \begin{equation}
 K(t,x,x) \sim t^{-d/2} + O(t^\infty).
 \label{Kloc}\end{equation}
 The local cylinder expansion, therefore, has the structure
 \begin{equation}
 T(t,x,x) \sim t^{-d} + {\bf E}(x)t + \cdots\,.
 \label{Tloc}\end{equation}
Let $T_{00}(t,x)$ be the energy density, regularized by an 
exponential cutoff as discussed below (\ref{bilinear}).
The relation between $T_{00}(t,x)$ and $T(t,x,x)$ is less 
straightforward than that between their global counterparts, 
 $E(t)$ and $T(t)$.
 The analogue of (\ref{Ereg}) yields only the terms in (\ref{T00})
 involving time derivatives.  The other terms can be obtained 
\cite{naples} by applying appropriate spatial partial derivatives 
to the integral kernel of the operator 
 \begin{equation}
 \frac1{\sqrt{H}} e^{-t\sqrt{H}}, 
 \label{othercylop}\end{equation}
 which has not been as extensively studied as the cylinder operator 
but has a very similar nature.
 (Note that the original cylinder kernel is the time derivative of 
this new one.)
 For the qualitative considerations of the present paper it is not 
necessary to delve further into the details; it is not misleading 
to think of $T_{00}(t,x)$ as ``essentially'' the $t$-derivative of 
 $T(t,x,x)$.

 This is as far as the heat kernel can take us.
 To find ${\bf E}(x)$ one must do independent calculations.
 But the literature, notably [\onlinecite{DC,KCD,RS}],
 provides considerable information about the behavior of 
the renormalized $T_{00}(x)$ as $x$ approaches a boundary.
Our task now is to compare these expansions in distance from the 
boundary (which we once again denote by $x$)
 to the expansions (in~$t$) of the regularized total energy,
the derivatives of (\ref{T1})--(\ref{T3}).
 The point is that the finite, renormalized energy density 
generally is not integrable up to the boundary, and if such
divergent terms are taken seriously (and not compensated by 
postulated terms concentrated exactly on the boundary \cite{KCD}, 
whose discussion we postpone to Sec.~\ref{sec:implic}),
 then for consistency they must somehow be reflected in 
terms of $E(t)$ that become {\em infinite\/} in the limit $t\to0$ 
and hence would be removed in a conventional renormalization of the 
energy.

 \subsection{Dimension 1}

 The energy density found by Romeo and Saharian \cite{RS} near a 
Robin endpoint (with $\xi=\frac14$)
 can be correlated with the derivative of (\ref{T1}) 
as follows:
 \begin{equation}
 \begin{array}{rcccccccccl}
 T_{00}& \sim & \infty & + &\displaystyle\frac S{x^2} &+&
\displaystyle\frac\gamma x&+&
 \gamma^2 \ln x &+& \mbox{finite} \\
&&\downarrow &&\mathop\downarrow\limits_? &&
  \downarrow&\searrow& \downarrow &\swarrow\!\!\!\!\!\!\searrow&\\
 E&\sim &\displaystyle\frac V{t^2} & & &+& \gamma\ln t &+&
 ({\bf E} +\gamma) &+& \gamma^2 t\; + \;\cdots \\
 \end{array}\label{D1}\end{equation}
The leading $\infty$ in $T_{00}$ is the universal, $x$-independent 
 formal vacuum energy of infinite empty flat space, which everyone 
agrees should be discarded in renormalization.
 It is included here because it clearly corresponds directly to the 
 volume divergence in $E$.
 The arrows indicate how other terms in the two series might be 
related, on the basis of their dependence on~$\gamma$.
 (Whether all these arrows represent genuinely existing 
relationships remains to be investigated by a thorough and rigorous 
calculation of $T_{00}(t,x,x)$, followed by an integration over $x$ 
with $t$ finite but approaching~$0$.  That project is beyond our 
present scope.)
 The most striking observation is that the ``surface'' term in the 
density has completely disappeared from the total energy.
 (Recall that the corresponding term in (\ref{T1}) was destroyed by 
the differentiation.)
 The $1/x$ term in the density might be held responsible for 
the $\,\ln t$ term in the energy, as well as part of the finite 
term.
The $\,\ln x$ term is integrable, and its integral is presumably 
included in ${\bf E}$.
 Finally, we note that all the explicitly $\gamma$-dependent terms 
in~$E$ will be independent of the length of the cavity, so they 
will not contribute to the Casimir force between the endpoints
 (usually considered the only quantity of experimental relevance).
 The force must come entirely from ${\bf E}$.

    \subsection{Dimension 2}
From now on, for brevity, the Robin function, $\gamma$, will not be 
explicitly indicated in the displays (\ref{D2}) and (\ref{D3}), 
because it enters in the same way as the curvature,~$\kappa$.
I do not know a reference for the dependence of $T_{00}(x)$ 
upon~$\kappa$ in dimension~$2$, 
 but it appears to be dictated by dimensional analysis and 
analogies with the Robin formula \cite{RS} and the 
 three-dimensional results \cite{DC}.
  \begin{equation}
 \begin{array}{rcccccccccccl}
 T_{00}& \sim & \infty & + &\displaystyle\frac S{x^3} &+&
 \displaystyle\frac\kappa {x^2}&+&
 \displaystyle\frac{\kappa^2}x&+&\kappa^3 \ln x &+& \mbox{finite} \\
&&\downarrow &&\downarrow&&\mathop\downarrow\limits_? &&
  \downarrow&\searrow& \downarrow &\swarrow\!\!\!\!\!\!\searrow&\\
 E&\sim &\displaystyle\frac V{t^3}&+&\displaystyle\frac S{t^2} & & &+& 
\kappa^2\ln t &+& ({\bf E} +\kappa^2) &+&
 \kappa^3 t\; + \;\cdots \\
 \end{array}\label{D2}\end{equation}
Most noteworthy here are the term $\kappa^2/x$ in the density and 
the corresponding term $\kappa^2 \ln t$ in the total energy.
 The analogous terms in dimension~3 are proportional to~$\kappa^3$.
When total energy calculations are done for both the inside and the 
outside of a boundary surface (a ``shell'' modeling a thin 
electrical conductor), it has often been noted that divergent 
contributions proportional to {\em odd\/} powers of~$\kappa$ will 
cancel, because $\kappa$ has opposite signs on the two sides.
 This fact gives rise to an important difference between even and 
odd dimensions.
 In even dimensions, the $\,\ln t$ term in $E$ is proportional to 
an even power of $\kappa$, so it survives even for thin shells.
 Logarithmic terms in $E(t)$ regularized by ultraviolet cutoff are 
precisely the divergences that persist in analytic regularization 
schemes, giving rise, under even the most liberal 
 interpretation \cite{BVW},
 at least to ambiguous finite terms in the renormalized energy that 
depend logarithmically on an arbitrary mass scale.
 In dimensional regularization, the $\kappa^d$ divergence shows up 
as a pole at each {\em even\/} value of~$d$ in the total Casimir 
energy associated with a spherical shell \cite{Milton}.
 Here we have seen in a more direct, elementary way that such a 
term is dictated by the known form of the heat kernel.

\subsection{Dimension 3}

 We are now in the domain of the classic results of Deutsch and 
Candelas \cite{DC} and Kennedy, Critchley, and Dowker 
\cite{KCD}, which also apply to the electromagnetic field.
  \begin{equation}
 \begin{array}{rcccccccccccl}
 T_{00}& \sim & \infty & + &\displaystyle\frac S{x^4} &+& 
\displaystyle\frac\kappa {x^3}&+&
 \displaystyle\frac{\kappa^2}{x^2}&+&
\displaystyle\frac{\kappa^3}x&+& O(\ln x) \\
&&\downarrow&&\downarrow &&\downarrow&&\mathop\downarrow\limits_? &&
  \downarrow&\searrow& \downarrow \\
 E&\sim &\displaystyle\frac V{t^4}&+&\displaystyle\frac S{t^3}&+&
\displaystyle\frac \kappa {t^2}
  & & &+& \kappa^3\ln t &+& ({\bf E} +\kappa^3) 
 \; + \;\cdots \\
 \end{array}\label{D3}\end{equation}
Let us review the litany of special features \cite{DC,Cand1}
 that make the Casimir force on
 a spherical conducting shell finite, even when calculated 
by integrating the energy density right up to the boundary:
 The volume term is removed by the local renormalization.
 The surface term vanishes for electromagnetism because of a 
cancellation between electric and magnetic contributions.
 The $\kappa$ and $\kappa^3$ terms cancel (between inside and 
outside) for any thin shell.
 That leaves~$\kappa^2$.
 In dimension~2, $\kappa$~is a symmetric matrix with two 
eigenvalues, $\kappa_1$~and~$\kappa_2$ 
 (the {\em principal curvatures\/}).
 There are thus two independent quadratic invariants to be 
integrated over $\partial M$, which may be chosen as 
 \begin{equation}
D \equiv \kappa_1\kappa_2\,, \qquad
 \Delta \equiv (\kappa_1-\kappa_2)^2.
    \label{quadinv}\end{equation}
For a sphere, $\Delta$ is identically zero.
 For any closed surface, $\int_{\partial M} D(x)\, dS$ is a 
topological invariant by the Gauss--Bonnet theorem,
 so its derivative with respect to the sphere's radius vanishes.
(More generally, for any bounded region~$M$, any quadratic integral 
of the type $\int_{\partial M} \kappa^2 \,dS$ is constant under a 
shape-preserving magnification of~$M$, even if $\Delta$ is 
involved.
 Note that this observation does not settle how the energy per unit 
length of an infinitely long cylinder depends on the radius, since 
the scaling of the area is different in that case.)
 As emphasized by Candelas \cite{DC,Cand1,Cand2},
 one would nevertheless expect infinite energy differences to 
result from shape-changing deformations of a closed surface, 
because of the $\Delta$ term.
 It is therefore significant that {\em no\/} term of the type
 $\kappa^2 /t$ can appear in the regularized total energy, $E(t)$, 
because of the structure of the traced cylinder expansion, 
 (\ref{cyl}) or (\ref{T3}).
 We return to this issue in the next section.

  \section{Implications}\label{sec:implic}

 The theory of vacuum energy in electromagnetism has increasing 
experimental verification and many indirect 
implications for observable phenomena.
 Nevertheless, the simplest and crispest theoretical models tend to 
be idealizations that are not directly testable.
 Perhaps for this reason, controversies persist about 
renormalization procedures and the physical significance of various 
divergences.  These disputes fall into two main categories.

 First, there is the broad question of whether ``logarithmic'' 
divergences are a sign of a bad model, or just an inherent 
renormalization ambiguity.
 Among the divergences that arise in energies regularized by 
 point-splitting or ultraviolet cutoffs, 
such as the second line of~(\ref{D3}),
 only those proportional to $\,\ln t$ survive \cite{Kay,BVW,CVZ} 
 in analytic 
regularization schemes, where they appear as poles at the physical 
values of the regularization parameters.
 The analytic methods therefore give  finite energies automatically
in many problems, without a need for a renormalization subtraction, 
but in some other problems are embarrassed by a surviving 
divergence whose removal introduces an unavoidable ambiguity 
parametrized by an arbitrary length or mass scale. 
 Some authors \cite{KCD,BVW} regard this as not at all a problem,
 while others regard the poles as ``truly disturbing'' 
[\onlinecite[p.~199]{Milton}] or at least puzzling.

 Second, and specific to Casimir calculations, is the question of 
whether the nonintegrable divergences in the {\em renormalized\/}
 energy density near boundaries must be taken seriously, even in 
cases when the renormalized total energy (or the observable force) 
is finite. 
In fact, how can such a divergence even be regarded as logically 
consistent with a finite total energy?
 Both Kennedy et al.\ \cite{KCD} and Milton and coworkers 
 \cite{MdRS,Milball,Milton} trust the calculations of finite total 
energies and forces more than the calculations of local energy 
densities. 
 Kennedy et al.\ argued that the infinities in $T_{00}(x)$ must be 
compensated by delta-function terms concentrated exactly on the 
boundary itself, so that the finite total energy predicted by 
 zeta-function calculations will result.
 (The boundary term in (\ref{totalE}) originated in this discussion 
in [\onlinecite{KCD}], without a clear explanation of how it could 
be relevant when $\xi=\frac14$ or $\gamma =0$ or~$\infty$.)
 Milton writes [\onlinecite[p.~237]{Milton}],
  ``\dots\ [T]he local vacuum-fluctuation energy density is to a 
  large extent meaningless.\dots\ [D]ivergences in the energy which 
  go like a power of the cutoff are probably unobservable, being 
  subsumed in the properties of matter.'' 

On the other side, Candelas and Deutsch \cite{DC,Cand1,Cand2}
 argued that the boundary infinities in the energy density are real 
(within the theory) and must be regarded as a defect of an overly 
idealized model.
 This position has recently been reexpressed by Graham, 
Jaffe, and coworkers \cite{GJ2,GJ3,GJ4,GO,OG}.
 The argument, stated in electromagnetic terms, is that no real 
material is perfectly conducting at arbitrarily high frequencies.
 When an idealized boundary is replaced by a more realistic model of 
the interaction of the quantized field with other matter,
 the infinite pileup of energy  near the boundary will become 
finite, but possibly very large and definitely physically real.
 A strong argument for this position is that the energy-momentum 
tensor acts as the source of the gravitational field, so a 
nonintegrable
singularity in it is physically unacceptable, even if the total 
energy comes out finite by virtue of a cancellation (as happens in
some scenarios involving thin shells).

The present paper does not resolve all these issues, but
 it casts 
some light on them and suggests some directions for future 
research that will clarify them further,
as indicated in the following remarks.

 \subsection{Thin shells; Robin divergences}

 Much has been made of the disappearance of some divergences by 
cancellation in the case of thin shell boundaries.
 The dimensional dependence (even vs.\ odd) of this effect is quite 
clear from (\ref{D1})--(\ref{D3}) and their antecedents.
 Those formulas also indicate analogous divergences specific to 
boundaries with nontrivial Robin coefficients, which were 
discovered in 
 [\onlinecite{RS}] but could have been anticipated from an 
understanding of the relation between vacuum energy and the heat 
kernel.  
 Note that there is no reason to expect any cancellation of 
exterior and interior Robin effects for a thin boundary.

 \subsection{Disappearing $x^{-2}$ terms; total derivatives;
conductors with anisotropic curvature}

A pervasive feature of (\ref{D1})--(\ref{D3}) is that terms in $T_{00}$ 
proportional to $x^{-2}$ ($x$ being distance to the boundary)
never have counterparts in the regularized total energy.
On grounds of dimensional analysis such terms would be expected to appear 
in $E(t)$ and be proportional to $t^{-1}$,
but the derivation of $E(t)$ from the heat and cylinder kernels shows that 
they are not there!
(The crucial fact is that terms proportional to $t^n \ln t$ occur in the 
cylinder kernel only if $n$ is odd and positive --- 
in particular, not for $n=0$.)
This absence of $t^{-1}$ terms was noted by Cognola et al.~\cite{CVZ},
and very recently by Bernasconi et al.~\cite{BGH}, who found that it 
persists when the exponential cutoff is replaced by a Gaussian one.

In dimension~$1$  the $x^{-2}$ term is precisely the one associated 
with nonconformal coupling ---
that is, the leading boundary divergence, which 
(in any dimension, but for a scalar field)
does not occur for 
$\xi=\xi_{\rm c}$ and arises for other values of $\xi$
from a term proportional to 
$(\xi-\xi_{\rm c}) \nabla\cdot(\phi\nabla\phi)$ in the 
formal energy density (cf.\ (\ref{T00}) and ensuing discussion).
Since this term is a total derivative, it {\em ought\/} to give $0$
when integrated. 
Effectively this seems to be true, since the renormalized total energy
is finite and {\em independent of $\xi$}.
As has been recognized (if not ``understood'') since 
[\onlinecite{DC}] and [\onlinecite{KCD}],
integration over~$x$ somehow does not commute with renormalization.
There is a simpler and completely understood analogous situation 
with the heat kernel:  
The local heat expansion (\ref{Kloc}) is nonuniform in~$x$, and hence
integrating it over~$M$ does not yield the correct trace expansion,
(\ref{K}).  The interesting terms in the latter come from a part
of $K(t,x,x)$ that concentrates entirely on the boundary in the limit
$t\to0$.
It is tempting to think of the boundary terms in $T_{00}$ 
(see  (\ref{D1})--(\ref{D3})) as the cylinder-kernel analogues of
these boundary terms in~$K$, but they are not:
they are nonvanishing even when $t=0$.
More likely, there is some other part of $T(t,x,x)$ that concentrates
{\em right on the boundary\/} in the limit and partly or wholly
compensates the integrals of the energy divergences {\em near\/}
the boundary.
This would corroborate, within the framework of ultraviolet 
regularization, the proposal of Kennedy et al.\ \cite{KCD}, made
in the framework of zeta-function regularization.
(It would not, however, solve the problem of a singular Einstein equation.)

We propose, therefore, a program to calculate $T(t,x,x)$ in various models
for {\em finite\/}~$t$ and to integrate it over~$x$, and then to examine
the limit $t\to0$. The expectation is that the results (including
contributions from the kernel of (\ref{othercylop}) if necessary) will be
fully consistent with the direct regularization and renormalization of the
total energy, and that the details of the process will demonstrate
precisely how the terms in the expansion of the renormalized $T_{00}(x)$
near the boundary are related, consistently, with the divergent terms in
$E(t)$. Some steps in this program can already be extracted from the
literature: Parentani \cite{Par} has made a similar observation in the
context of a uniformly accelerating boundary. Ford and Svaiter \cite{FS},
also with an eye to resolving the paradox of the total energy of the
nonconformally coupled scalar field, studied a statistically fluctuating
boundary. Their formulas seem mathematically identical to the results of
ultraviolet regularization with an arbitrary (not necessarily exponential)
cutoff function, with a fluctuation parameter in the place of~$t$. They
find (in dimension~$3$) that the nonconformal term of $T_{00}$ oscillates
from negative to positive and back to negative, in such a way its integral
is indeed zero. Finally, Graham et al. \cite{GJ3,GJ4,OG,GO} replace the
ideal boundary with an extended object modeled by a second quantized
field. Their results are very similar to those of [\onlinecite{FS}] in
dimension~$3$ and have one fewer oscillation in dimension~$2$. (Olum and
Graham \cite{OG} also point out that the boundary itself, as a physical
object, has an energy density of its own that can be expected to swamp the
energy induced in the scalar field; nevertheless, they conclude that the
{\em total\/} energy density remains negative in a small region displaced
slightly from the boundary.)
All the available results suggest that the cancellation is insensitive to 
the functional form of the ultraviolet cutoff, and also to its physical 
interpretation (fluctuating boundary, soft boundary, or mere mathematical 
device).

In dimensions greater than $1$ the leading boundary divergence 
associated with
a nonconformal stress tensor does not go as~$x^{-2}$,
so the additional boundary terms that render the regularized energy 
$\xi$-independent are not related 
(at least, not directly)
to the $x^{-2}$-versus-$t^{-1}$ phenomenon.
Instead, that phenomenon would appear to have some other implications.
In particular, in dimension~$3$ it indicates that $E$ contains no term 
proportional to $\Delta = (\kappa_1-\kappa_2)^2$.
This conclusion seems to be robust against changes in the form of the 
ultraviolet cutoff \cite{BGH}. 
But local density calculations indicated to Candelas and 
Deutsch \cite{DC,Cand1,Cand2} that that term would be not only large but
negative 
(persisting into the renormalized energy for a physical conductor),
thereby rendering a conducting foil unstable against wrinkling.
This may well be true for a physical, imperfect conductor.
However, in the framework of formal ultraviolet regularization of the 
idealized theory,  
our conjecture is that there is a compensating positive term
that concentrates right on the boundary as the regularization parameter 
is taken to~$0$, thereby removing the apparent conflict between
local and global calculations. 
Clearly, detailed calculations are needed to verify this 
conjecture and to establish its relevance to realistic conductors.
(Incidentally, the sharp-eyed modern reader may notice that the original 
argument
concerning the $\Delta$ term in Appendix~A of [\onlinecite{DC}] 
incorrectly assumed that it is
correlated with a term of order $\omega^{-1}$ in a formal 
expansion of the counting function~$\mu$.
Indeed, Candelas pointed out in a later paper \cite[Sec.~7.3]{Cand1}
that no such term exists 
in~$\mu$, but that its alleged effects can still arise in the integrated
contribution of the oscillatory part of the eigenvalue distribution.
In the terminology of our Sec.~\ref{sec:notat},
the relevant terms in the cylinder kernel and in $T_{00}$ are associated
with constants of integration in Riesz means rather than with the local
behavior of $\mu'(\lambda)$, or even $\mu(\lambda)$, at large~$\lambda$.
Therefore, this tangential technical issue in no way invalidates the 
physical conclusions of [\onlinecite{DC}].)

 \subsection{Genuine surface effects}

 On the other hand, almost certainly there are some real
  near-boundary energies.
 In (\ref{D1})--(\ref{D3}) one observes terms of order $x^{-1}$ 
 in the renormalized energy density, which appear to be related to 
the divergent terms of order $\,\ln t$ in the regularized total 
energy (although explicit calculations, of the sort described 
above, should be performed to confirm this).  As previously 
mentioned,  terms of the latter type are agreed to yield 
terms with inherently ambiguous coefficients in the renormalized 
energy. 
 Also, terms of order $\,\ln x$ in $T_{00}$ are integrable and 
presumably contribute part of the finite term in~$E$.

 \subsection{Nonrelativistic Casimir energy?}

 On p.~30 of [\onlinecite{Milton}], Milton raises the question of 
why, mathematically, there is no Casimir energy in nonrelativistic 
quantum theory.
(Physically, such energy can't exist because the relativistic 
effect vanishes exponentially with large field mass.)
 He mimics the relativistic zeta-function calculation 
--- for parallel plates with Dirichlet conditions --- and 
 finds an expression containing a gamma function in the 
denominator, whose pole forces the quantity to vanish in the 
relevant limit.
The same conclusion can be reached more transparently from our 
   point of view.
 The formal energy is
 \[ E=\frac12 \sum_n {\omega_n{}\!^2 \over 2m}\,,\]
 which can be regularized in analogy with (\ref{Ereg}) as
 \[ \frac1{4m} \, \frac{d^2}{d t^2}
 \sum_n e^{-t\omega_n}. \]
 But the sum here is simply $T(t)$, which has, as in
  (\ref{D1})--(\ref{D3}), the schematic form
 \[ T(t) \sim \mbox{divergent terms} + {\bf E} t
 + (\kappa^{d+1} + \cdots+ \gamma^{d+1}) t^2 + {\bf F}t^3 
 +\cdots, \]
 the coefficient of the $t^2$ term being dictated by the 
 heat-kernel expansion.
 Therefore, in the renormalized energy one gets
 \[  \left. {d^2 T \over d t^2}\right|_{t=0} =0 \]
because $\kappa$ and $\gamma$ are zero in the system studied. 

\bigskip In summary, what we have shown is that various results for
specific models that previously emerged from forests of gamma and Riemann
zeta functions now are easy consequences of the known general forms of the
heat and cylinder expansions.  There are, to be sure, enough gamma
functions in [\onlinecite{riesz}] and Sec.~\ref{sec:notat} to satisfy
anyone, but they are handled once and for all in the general theory,
rather than arising in each concrete calculation out of some physically
opaque analytic continuation. With respect to the controversies over
global surface-energy renormalization versus local energy-density
calculations, we have not taken a position squarely on either side.  
Rather, we have presented evidence that each side is correct in certain
respects. In particular, a mechanism of partial surface cancellation has
been pointed out that shows promise of reconciling the two points of view.  
For the leading boundary divergence of the scalar field, this mechanism
has been fairly well established by other researchers \cite{Par,FS,GO};
for the $x^{-2}$ divergence, it is conjectured by the present author
as the only visible way of restoring consistency between the mathematical 
structures of the local and global asymptotic expansions. 

\goodbreak
\begin{acknowledgments}
This material was first presented at the Workshop on Casimir Forces 
at the Harvard--Smithsonian Center for Astrophysics; 
 I am grateful 
to the organizers for the invitation 
 and to many of the 
participants, especially Larry Ford, Nami Svaiter, and Kim Milton, 
for comments and bibliographical information. 
 The manuscript was 
written during a semester at MSRI, 
 and I acknowledge the financial 
generosity of both that Institute and my home department that made 
that stay possible. 
 Joel Bondurant aided in the study of the energy density and its 
boundary term in their dependence on $\xi$ and~$\gamma$. 

\end{acknowledgments}

\goodbreak

\end{document}